\documentclass[aip, twocolumn, amsmath, amssymb, reprint]{revtex4-2}

\usepackage{graphicx}
\usepackage{dcolumn}
\usepackage{bm}
\usepackage{color}

\usepackage[utf8]{inputenc}
\usepackage[T1]{fontenc}
\usepackage{mathptmx}
\usepackage{etoolbox}
\usepackage{orcidlink}

\makeatletter
\def\@email#1#2{%
 \endgroup
 \patchcmd{\titleblock@produce}
  {\frontmatter@RRAPformat}
  {\frontmatter@RRAPformat{\produce@RRAP{*#1\href{mailto:#2}{#2}}}\frontmatter@RRAPformat}
  {}{}
}%
\makeatother

\begin{document}

\preprint{AIP/123-QED}

\title{Beyond the standard model of topological Josephson junctions:\\From crystalline anisotropy to finite-size and diode effects}
\author{Bar{\i}\c{s} Pekerten \orcidlink{ 0000-0002-5794-9706}}
\email[\textcolor{red}{Corresponding author: }]{barispek@buffalo.edu}
\affiliation{Department of Physics, University at Buffalo, State University of New York, Buffalo, New York 14260, USA}
\author{David Brand{\~a}o \orcidlink{0000-0002-5772-1832}}
\affiliation{Department of Physics, University at Buffalo, State University of New York, Buffalo, New York 14260, USA}
\author{Bailey Bussiere \orcidlink{0009-0009-7117-5784}}
\affiliation{Department of Physics, University at Buffalo, State University of New York, Buffalo, New York 14260, USA}
\author{David Monroe \orcidlink{0000-0002-4640-8912}}
\affiliation{Department of Physics, University at Buffalo, State University of New York, Buffalo, New York 14260, USA}
\author{Tong Zhou \orcidlink{0000-0003-4588-5263}}
\affiliation{Department of Physics, University at Buffalo, State University of New York, Buffalo, New York 14260, USA}
\affiliation{Eastern Institute for Advanced Study, Eastern Institute of Technology, Ningbo, Zhejiang 315200, China}
\author{Jong E. Han \orcidlink{0000-0002-5518-2986}}
\affiliation{Department of Physics, University at Buffalo, State University of New York, Buffalo, New York 14260, USA}
\author{Javad Shabani \orcidlink{0000-0002-0812-2809}}
\affiliation{Center for Quantum Phenomena, Department of Physics, New York University, New York, New York 10003, USA}
\author{Alex Matos-Abiague}
\affiliation{Department of Physics and Astronomy, Wayne State University, Detroit, MI 48201, USA}
\author{Igor \v{Z}uti\'{c} \orcidlink{0000-0003-2485-226X}}
\email[\textcolor{red}{Corresponding author: }]{zigor@buffalo.edu}
\affiliation{Department of Physics, University at Buffalo, State University of New York, Buffalo, New York 14260, USA}

\date{\today}

\begin{abstract}
A planar Josephson junction is a versatile platform to realize topological superconductivity over a large parameter space and host Majorana bound states.  With a change in Zeeman field, this system undergoes a transition from trivial to topological superconductivity accompanied by a jump in the superconducting phase difference between the two superconductors. A standard model of these Josephson junctions, which can be fabricated to have a nearly perfect interfacial transparency, predicts a simple universal behavior. In that model, at the same value of Zeeman field for the topological transition, there is a $\pi$ phase jump and a minimum in the critical superconducting current, while applying a controllable phase difference yields a diamond-shaped topological region as a function of that phase difference and a Zeeman field. In contrast, even for a perfect interfacial transparency, we find a much richer and nonuniversal behavior as the width of the superconductor is varied or the Dresselhaus spin-orbit coupling is considered. The Zeeman field for the phase jump, not necessarily $\pi$, is different from the value for the minimum critical current, while there is a strong deviation from the diamond-like topological region. These Josephson junctions show a striking example of a nonreciprocal transport and superconducting diode effect, revealing the importance of our findings not only for topological superconductivity and  fault-tolerant quantum computing, but also for superconducting spintronics. 
\end{abstract}

\maketitle

Josephson junctions (JJs), with two superconductors (S) separated by a nonsuperconducting (normal) region (N) provide a fascinating manifestation of proximity effects, where a given material is transformed by its neighbors.\cite{Zutic2019:MT,Amundsen2022:RMP} The superconductivity leaking from the two S regions can merge, establishing a relative phase difference, $\phi$, transforming the whole N region into a superconductor.\cite{Tafuri:2019} Decades before Josephson's prediction,\cite{Josephson1962:PL} this behavior was realized experimentally.\cite{Holm1932:ZP,Zutic2019:MT} Even in conventional materials, through proximity effects, JJs reveal an exotic emergent behavior, absent in any of their constituent regions. With the interplay between spin-orbit coupling (SOC) and Zeeman field (applied or due to exchange coupling), the proximity-induced superconductivity could be transformed from a conventional spin-singlet  into spin-triplet $p$-wave superconductivity. This system could support topological superconductivity over an enhanced parameter range by the applied relative phase difference.\cite{Fu2008:PRL,Keselman2013:PRL,vanHeck2014:PRB,Kotetes2015:PRB,*Kotetes2020:PRB,Hell2017:PRL,Pientka2017:PRX,Zhou2020:PRL}

After Kitaev's pioneering work on spin-triplet topological superconductivity and Majorana bound states (MBS),\cite{Kitaev2001:PU,Kitaev2003:AP} an early work suggested planar JJs as materials realization that supports non-Abelian properties.\cite{Fu2008:PRL} However, most of the subsequent studies have focused on proximitized semiconductor nanowires\cite{Lutchyn2010:PRL,Oreg2010:PRL,Rokhinson2012:NP,Leijnse2012:SST,Aasen2016:PRX,Laubscher2021:JAP,Marra2022:JAP} and MBS detection through the quantized zero-bias conductance peak,\cite{Sengupta2001:PRB,Mourik2012:S,Lee2012:PRL,Das2012:NP} which can also arise from extrinsic effects.\cite{Chen2019:PRL,DasSarma2021:PRB,Yu2021:NP} Experiments have realized a high-transparency epitaxially grown planar JJs with proximity-induced superconductivity from Al into the InAs two-dimensional electron gas (2DEG).\cite{Shabani2016:PRB} This advance has stimulated important theoretical efforts recognizing that topological superconductivity can be supported over a larger parameter range than in semiconductor nanowires, while also providing phase-sensitive signatures,\cite{Hell2017:PRL,Pientka2017:PRX,Scharf2019:PRB} as well other opportunities in 2D platforms.\cite{Virtanen2018:PRB,Zhou2022:NC,Zhou2020:PRL,Setiawan2019:PRB,Fatin2016:PRL,Matos-Abiague2017:SSC,Paudel2021:PRB,Pankratova2020:PRX,Gungordu2022:JAP}

\begin{figure}[t]
\centering
\includegraphics*[width=\columnwidth]{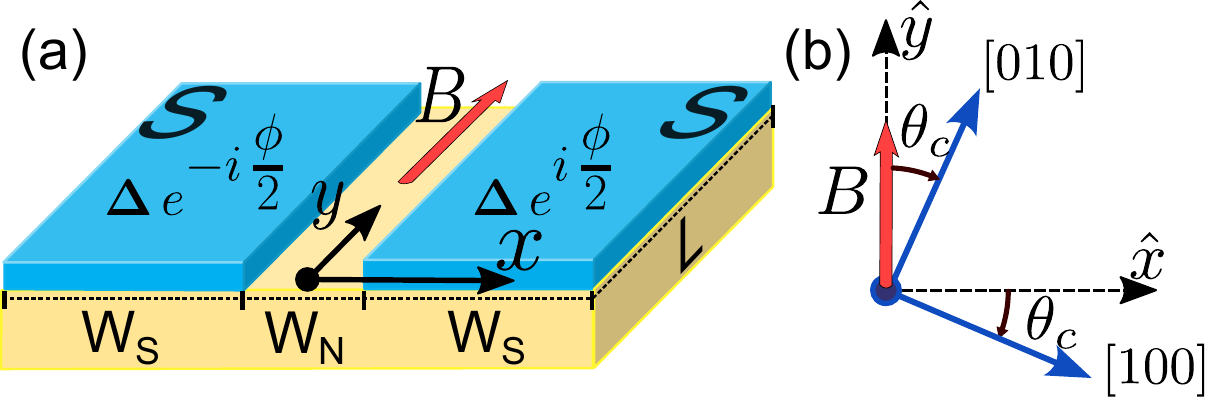}
\caption{(a) Planar JJ formed by two superconducting (S) regions with superconducting gap $\Delta$ and phase difference $\phi$ covering a 2DEG normal (N) region. Coordinate axes and JJ dimensions are indicated. The supercurrent $I$ flows along $\pm\hat{x}$, perpendicular to the applied magnetic field $\mathbf{B}$. (b) The misalignment of the crystallographic direction $[010]$ and $\mathbf{B}$ is represented by the angle $\theta_c$ which, with Dresselhaus SOC, 
leads to crystalline anisotropy.}
\label{fig:System} 
\end{figure}

Experiments on planar JJs,\cite{Fornieri2019:N,Ren2019:N,Dartiailh2021:PRL,Banerjee2023:PRB,Banerjee2023:PRL,Li2024:NL} including detecting the phase signature of topological superconductivity\cite{Dartiailh2021:PRL} and a superconducting diode effect,\cite{Dartiailh2021:PRL,Baumgartner2022:NN,Nadeem2023:NRP,Lotfizadeh2024:CP} reveal a much richer physics than the standard model. Such a model predicts a universal behavior for highly-transparent interfaces, with the diamond-like topological region as a function of $\phi$ and in-plane Zeeman field.\cite{Pientka2017:PRX} In this work, we provide a more general description of planar JJs for the geometry illustrated in Fig.~\ref{fig:System}(a), where $W_S$ ($W_N$) the width of the S (N, uncovered 2DEG) region and $L$ is the length. In Fig.~\ref{fig:System}(b), the presence of Dresselhaus SOC in the 2DEG\cite{Zutic2004:RMP,Fabian2007:APS} and a related crystalline anisotropy is represented by the angle $\theta_c$.\cite{Pakizer2021:PRR,Pakizer2021:PRB,Pekerten2022:PRB} We show that a finite $W_S$ and Dresselhaus SOC each strongly modify the properties of planar JJs implied by the standard model.

We solve the discretized Bogoliubov-de Gennes (BdG) equation using a finite-difference method for BdG Hamiltonian, $H_\text{BdG}$, in the Nambu basis $( \psi_\uparrow, \psi_\downarrow, \psi^\dagger_\downarrow, -\psi^\dagger_\uparrow)$ with $\psi$ ($\psi^\dagger$) destruction (creation) operators of the given spin\cite{Pakizer2021:PRR,Pekerten2022:PRB}
\begin{align}
H_\text{BdG} & = \left[\frac{\mathbf{p}^2}{2m^\ast} - \mu\left(x,y\right) + 
H_\text{SO}\right]
\tau_z - \frac{g^\ast\mu_B}{2}\mathbf{B}\cdot\boldsymbol{\sigma}\nonumber\\      
	&\quad  + \Delta\left(x,y\right)\tau_+ + \Delta^\ast\left(x,y\right)\tau_-\;,
\label{eq:H_BdG}
\end{align}
where $H_\text{SO}$ contains Rashba and Dresselhaus SOC terms
\begin{eqnarray}
H_\text{SO}&=& \frac{\alpha}{\hbar}(p_y\sigma_x-p_x\sigma_y)  +\frac{\beta}{\hbar} \big[(p_x\sigma_x-p_y\sigma_y)\cos{2\theta_c}\nonumber \\
&-& (p_x\sigma_y+p_y\sigma_x) \sin{2\theta_c} \big]. 
\label{eq:SOC}
\end{eqnarray}
Here, ${\bf p}$ is the momentum, $\mu\left(x,y\right)$ is the chemical potential, $\mathbf{B} = B\,\hat{y}$ is the in-plane magnetic field, $\mu_B$ is the Bohr magneton, while $m^\ast$ and $g^\ast$  are the effective mass and $g$-factor. The proximity-induced \textit{s}-wave superconducting gap term is $\Delta(x,y) = \Delta_0 \, f_\mathrm{BCS}(B, T) \, \Theta(|x-W_N/2|)\,\exp{[i\,\phi(x, y)/2}]$, where $\Theta$ is the  step function and $\phi(x, y) = \phi_0 \, \mathrm{sgn}{(x)}$, with $\phi_0$ the phase difference across the JJ, while $\Delta_0$ is real and uniform in the S regions. The usual BCS dependence of $\Delta(x,y)$ on $B$ and temperature, $T$, is given by $f_\mathrm{BCS}(B, T)$ with $f_\mathrm{BCS}(B=0, T=0)\,=\,1$.\cite{Tinkham:1996} $\alpha$ ($\beta$) is the Rashba (Dresselhaus) SOC strength. We consider that  $y$ direction is translationally invariant in Eq.~(\ref{eq:H_BdG}), such that wavevector $k_y = p_y\,/\,\hbar$ is a good quantum number, and numerically obtain the energy spectrum and topological properties of its discretized version.\cite{Groth2014:NJP,Pekerten2022:PRB}

Planar JJs can be phase biased; for example, $\phi$ can be externally set by embedding a JJ in a superconducting loop threading the magnetic flux through the loop. Alternatively, a phase-unbiased JJ attains its ground state value, $\phi_{GS}$, at which the free energy of the system, $F(\phi, B)$,  is minimized and satisfies $\partial_\phi F(\phi_{GS}, B)=0$. In phase-unbiased topological JJs the system self-tunes into a topological state as the in-plane $B$ is varied by causing a phase jump in $\phi_{GS}\sim\pi$. We obtain\cite{Tinkham:1996}
\begin{align}\label{eq:FE}
F(\phi, B) &=-k_B T \sum_{E_n, k_y} \mathrm{ln}\left[2\,\mathrm{cosh} \left( E_n(k_y, B)/2 k_B T \right) \right],
\end{align}
from the calculated spectrum $\{E_n(k_y, B)\}$, where $k_B$ is the Boltzmann constant. The current-phase relation (CPR), $I(\phi, B) = (2e/\hbar) \, \partial F/\partial\phi$, is then 
\begin{align}
I(\phi, B) &= -\frac{e}{\hbar}\, \sum_{E_n, k_y}\,\frac{\partial E_n(k_y)}{\partial \phi} \, \mathrm{tanh}\left( \frac{E_n(k_y, B)}{2 k_B T} \right).
\label{eq:CPR} 
\end{align}
The extrema of $F(\phi, B)$ corresponds to the zeros of the supercurrent, $I(\phi, B) \propto \partial_\phi F(\phi, B) = 0$, hence $\phi_{GS}$ follows one of the zeros of the CPR.\cite{Pekerten2022:PRB} The forward (reverse) critical current, $I_c^\pm(B)$, is  $I_c^\pm(B) = \left|\min_\phi\,(\max_\phi) \,I(\phi, B)\right|$, and the total critical current is $I_c(B) = \max(I_c^+, I_c^-)$.We normalize critical currents, $\widetilde{I\,}_c(B) \equiv I_c(B)\,/\,I_{c,\mathrm{max}}$, $\widetilde{I\,}_c^\pm(B) \equiv I_c^\pm(B)\,/\,I_{c,\mathrm{max}}$, to the maximum critical current,  $I_{c,\mathrm{max}} \equiv \max_B I_c(B)$.  Similarly, $I_{c, \mathrm{min}}$, and  $\widetilde{I\,}_{c, \mathrm{min}}\equiv I_{c, \mathrm{min}}\,/\,I_{c,\mathrm{max}}$, denote the critical current at its dip near the topological transition and its normalized value.

The class D topological charge, $Q_D$,\cite{Altland1997:PRB,Kitaev2001:PU,Schnyder2009:AIP,Ryu2010:NJP,Ghosh2010:PRB,Tewari2012:PRL} for the JJ  is 
\begin{align}
Q_D &= {\rm sgn}\left[{\rm Pf}\{H(k_y = \pi)\sigma_y\tau_y\}/{\rm Pf}\{H(k_y = 0)\sigma_y\tau_y\}\right],
\label{eq:Q_D}
\end{align}
where ${\rm Pf}\{...\}$ denotes the Pfaffian, with $Q_D=1\, (-1)$ determining the system in the trivial (topological) phase. We employ experimentally relevant parameters for Al/InAs planar JJs,\cite{Dartiailh2021:PRL} with $\Delta_0=0.23$meV, $m^\ast=0.027m_0$, where $m_0$ is the bare electron rest mass, and $g^\ast=10$. We parametrize the strengths of Rashba and Dresselhaus SOC through an overall strength, $\lambda_\mathrm{SO}$, and an angle, $\theta_\mathrm{SO}$, defined as
\begin{equation}
\lambda_\mathrm{SO}\equiv\sqrt{\alpha^2+\beta^2}\;,\;\theta_\mathrm{SO}\equiv{\rm arccot}(\alpha/\beta)\;,
\label{param}
\end{equation}
with $\alpha=\lambda_\mathrm{SO}\,\cos\theta_\mathrm{SO}$ and $\beta=\lambda_\mathrm{SO}\,\sin\theta_\mathrm{SO}$.\cite{Pekerten2022:PRB,Pakizer2021:PRR} We take $\mu(x, y) = \mu_0 - \epsilon(\lambda_\mathrm{SO}, \theta_c)$ constant throughout the system, with $\mu_0=1\;$meV and  $\epsilon( \lambda_\mathrm{SO}, \theta_c) = (2 m^*\lambda_\mathrm{SO}^2/\hbar^2) \, \left[1+(\sin{2\theta_c})^2 \right]$ denoting the bottom of the single particle energy band.

\begin{figure}[t]
\centering
\includegraphics*[width=\columnwidth]{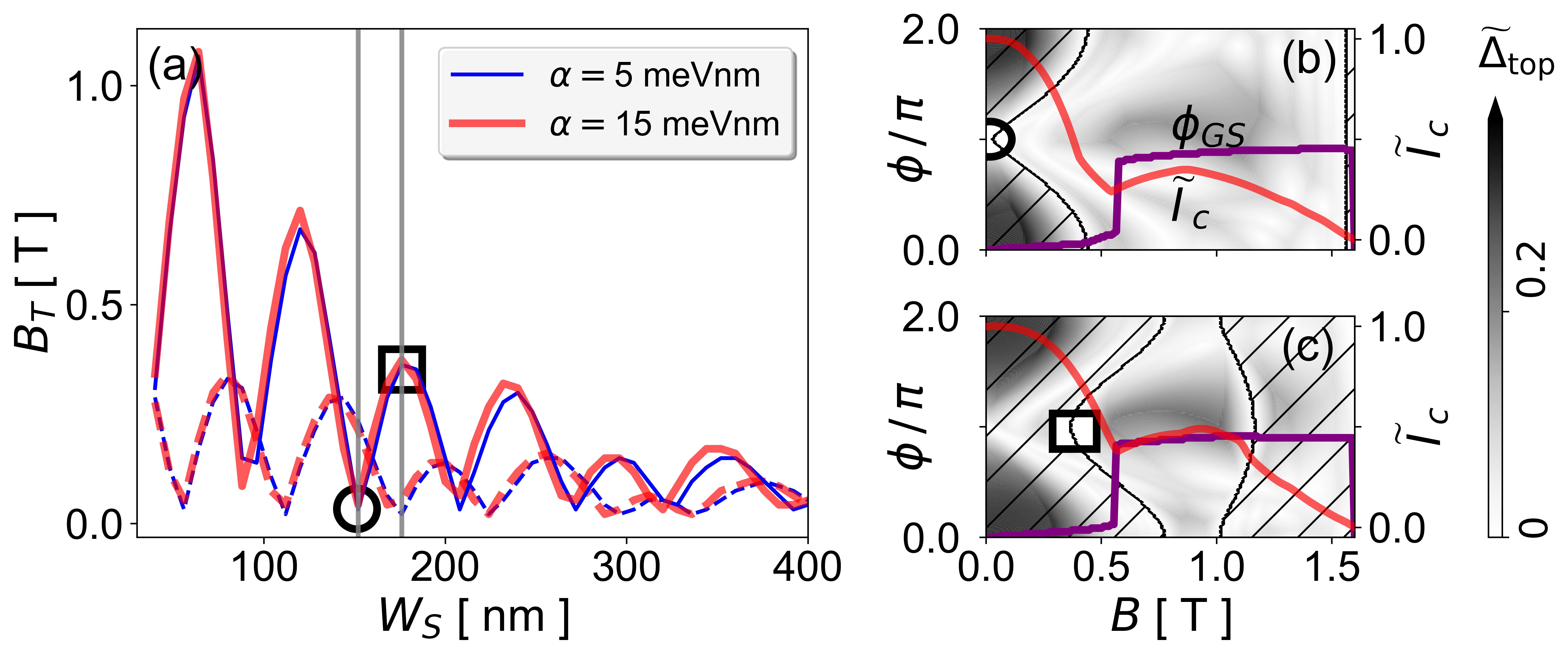}
\caption{Finite size effects with $W_N = 160,\,600\;$nm (solid, dashed) and $\alpha=5,\,15\;$meVnm (blue, red). (a) Oscillating minimum topological Zeeman field $B_T(W_S)$ at $\phi=\pi$. (b), (c) Topological phase diagram at a dip, peak in $B_T$ at $W_S=152,\,176\;$nm [vertical lines and circle, rectangle in (a)] showing topological charge $Q_D(\phi, B)=1,\,-1$ (hatched, unhatched) and normalized topological gap $\widetilde{\Delta}_\mathrm{top}(\phi, 
B)$ (grayscale), with superimposed ground state phase $\phi_\mathrm{GS}(B)$ (violet, left axis) and normalized critical current $\widetilde{I}_c^\pm(B)$ (red, right axis).}
\label{fig:BT_Oscillations} 
\end{figure}

The standard model of topological JJs  provides important ideas for topological superconductivity and its phase control.\cite{Pientka2017:PRX} By assuming Rashba SOC and a perfect transparency,  $\tau=1$,\cite{Pientka2017:PRX} at N/S interfaces, which is realized in Al/InAs JJs with $\tau=0.98$,\cite{Kjaergaard2017:PRA,Nichele2020:PRL} that model leads to a universal behavior for topological superconductivity. The topological region with $Q_D(\phi, B)=-1$ has a diamond shape,\cite{Pientka2017:PRX} formed by closing of a topological gap for $\phi=\pi$, at $B=0$ and $B^*=\pi\sqrt{2m^*\mu}/(g^*\mu_B W_N)$. For $\phi=0,2\pi$ and $Q_D=1$ at any $B$, the topological region shrinks to zero. For a phase-unbiased JJ, at $B^*/4$, a topological transition is accompanied by: (i) dip (local minimum) in  $\widetilde{I}_c$ and (ii) $0-\pi$ jump in $\phi_{GS}$.

With the same assumption of the perfect N/S transparency, in Fig.~\ref{fig:BT_Oscillations} we find a striking difference from the predicted universal $B^*\propto 1/W_N$ behavior and a diamond-shape topological region. Finite-size effects in $W_S$ lead to the oscillation of the minimum in-plane Zeeman field for the transition into a topological state, $B_T$, at $\phi = \pi$, where $Q_D$ switches from $1$ to $-1$.\cite{Pientka2017:PRX,Tewari2012:PRL} In Fig.~\ref{fig:BT_Oscillations}(a), for the considered N-region widths, $W_N=160$ $(600)\;$nm, solid (dashed) line, as well the Rashba SOC strengths, $\alpha=5$ $(15)\;$meVnm, blue (red) line, used throughout our calculations, we see strong oscillations in $B_T(W_S)$. These oscillations also deform the diamond-shaped topological region. This means that, at $\phi = \pi$, a suitable choice of $W_S$, reducing $B_T$, can relax the requirements for the onset of topological superconductivity in various materials.

These decaying oscillations in $B_T$, defined at $\phi=\pi$, are also translated to changes in the whole topological region. For a fixed $W_N=160\;$nm, even a small change in $W_S=152, 176\;$nm [vertical lines mark the corresponding dip and peak in $B_T$ from Fig.~\ref{fig:BT_Oscillations}(a)], leads to a large change in the related size and shape of the topological region in Figs.~\ref{fig:BT_Oscillations}(b) and \ref{fig:BT_Oscillations}(c), respectively. For example, in Fig.~\ref{fig:BT_Oscillations}(b), we see a topological phase diagram, bounded by a hatched region and $Q_D=1$. The topological region is large and has a small $B_T$, even though there is an overall  distortion of the diamond shape. However, while the topological region with $Q_D=-1$ implies that MBS can be supported, it is also important to know  the corresponding (normalized) topological gap,\cite{Hell2017:PRL,Pakizer2021:PRR,Pakizer2021:PRB,Pekerten2022:PRB} $\widetilde{\Delta}_\mathrm{top}\equiv \Delta_\mathrm{top}\,/\,\Delta_0$, shown as the gray scale background. Specifically, MBS are not well protected in the regions with a small $\Delta_\mathrm{top}(\phi, B) \equiv \min\left(\{E_n(k_y, \phi, B)>0\}\right)$, where two end-MBS hybridize with each other and move away from zero energy. In contrast, for a slightly wider $W_S$ in Fig.~\ref{fig:BT_Oscillations}(c), we see a much larger $B_T$ and a smaller topological region than in Fig.~\ref{fig:BT_Oscillations}(b).

In previous studies,\cite{Pientka2017:PRX,Scharf2019:PRB,Setiawan2019b:PRB} changes from the diamond-shape topological region were attributed to the imperfect transparency which leads to the normal reflection at N/S interfaces, not just Andreev reflections (responsible for the formation of Andreev bounds states in JJs). However, even for perfect N/S interfaces, we note that with a finite $W_S$ there are normal reflections at the ends of the S regions. The quasiparticle wavefunctions entering the S regions are reflected back from their ends into the N region and contribute to the normal reflection at the N/S interfaces. The $B_T(W_S)$ oscillations in Fig.~\ref{fig:BT_Oscillations}(a) can be understood from the resonant transmission condition, in which the quasiparticle wavefunction forms a standing wave in the S region.\cite{Tinkham:1996} This can lead to the transition into the topological phase at a smaller in-plane $B$ and offer a wider topological region for experimentally relevant parameters. The decay of the $B_T(W_S)$ oscillations is consistent with the semiclassical picture of the  decaying quasiparticle wavefunction in the S region of width $W_S$, until it reenters the N region after a travel of $\sim 2W_S$ in the superconductor. With $W_S \rightarrow \infty$, this source of the normal reflection vanishes and thus minimizes $B_T$. 

For chosen parameters in Figs.~\ref{fig:BT_Oscillations}(b) and ~\ref{fig:BT_Oscillations}(c), a closer look suggest other deviations from the standard picture. The dip in $\widetilde{I}_c$  and $0-\pi$ jump in $\phi_{GS}$ no longer take place at the same value of $B$, which can be much more pronounced in other cases.\cite{Dartiailh2021:PRL,Pekerten2022:PRB} The jump in $\phi_{GS}$ is generally different from $\pi$,\cite{Pekerten2024:P} the size of the jump decreases with $\alpha$. Together, this suggests that, even for a perfect N/S transparency,  there is a considerably richer behavior for topological JJs than commonly expected, warranting our further analysis and identifying opportunities to facilitate experimental implementation of JJs.

\begin{figure}[t]
\centering
\includegraphics*[width=\columnwidth]{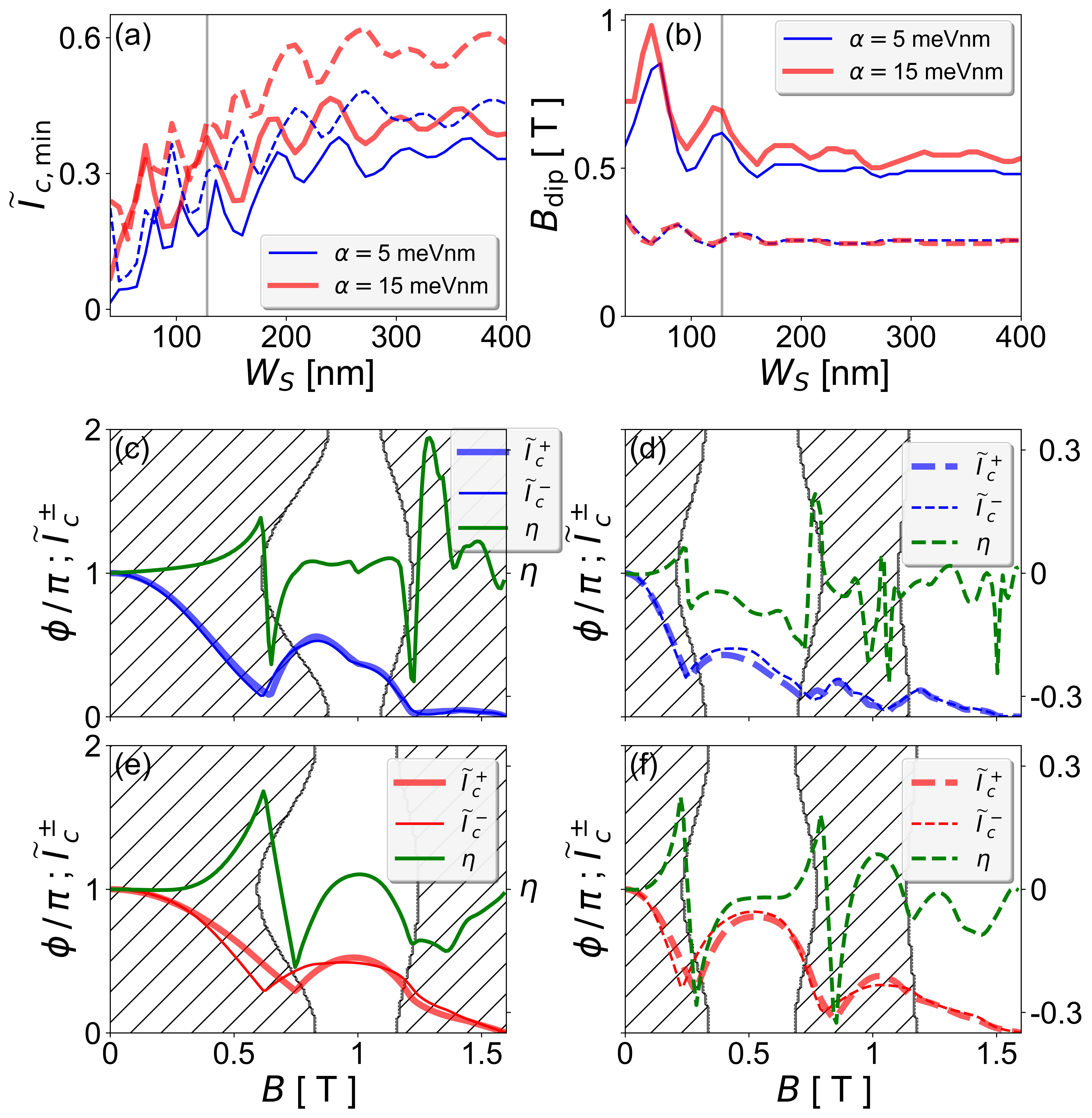}
\caption{Critical current dip near the first topological transition and superconducting diode effect; solid/dashed lines and colors follow the convention introduced in Fig.~\ref{fig:BT_Oscillations} for $W_N$ and $\alpha$. (a) Oscillating critical current at dip, $\widetilde{I\,}_{c, \mathrm{min}}(W_S)$. (b) Zeeman field at dip, $B_\mathrm{dip}(W_S)$. (c) - (f) Diode effect for $W_S=128\;$nm [gray line in (a), (b)]: unequal forward, reverse normalized critical current $\widetilde{I\,}_c^\pm(B)$ (thick, thin line; left axis), diode effect efficiency $\eta(B)$ for $\alpha=15\;$meVnm (green, right axis), and topological charge $Q_D=1,\,-1$ (hatched, unhatched).}
\label{fig:IcminLoc} 
\end{figure}

With this motivation, we next investigate the size and location of the dip of $\widetilde{I\,}_c$, denoted by $\widetilde{I\,}_{c,\mathrm{min}}$ and $B_\mathrm{dip}$, respectively, and shown in Figs.~\ref{fig:IcminLoc}(a) and \ref{fig:IcminLoc}(b). This dip corresponds to the minimization of the singlet component of the finite-momentum wavefunction, which vanishes near the topological phase transition in the absence of normal reflection in the JJ.\cite{Pientka2017:PRX,Pakizer2021:PRR,Pekerten2022:PRB,Lotfizadeh2024:CP} At a fixed temperature, the magnitude of $\widetilde{I\,}_c$  is thus related to the  contribution of the normal reflection. Therefore, oscillations in $\widetilde{I\,}_c(W_S)$,  similar to those in Fig.~\ref{fig:BT_Oscillations}(a), are also  present in Fig.~\ref{fig:IcminLoc}(a). However, unlike only a weak dependence of the magnitude of $B_T(W_S)$ oscillations with $\alpha$ in Fig.~\ref{fig:BT_Oscillations}(a), the same change in the Rashba SOC strongly changes the magnitude of the $\widetilde{I\,}_c(W_S)$ in Figs.~\ref{fig:IcminLoc}(a), for both $W_N=160$ and $600\;$nm. Turning to the dip location in Fig.~\ref{fig:IcminLoc}(b), $B_\mathrm{dip}(W_S)$ shows decaying oscillations which are strongly suppressed for a larger $W_N$ and with increasing $W_S$. Since a finite $W_S$ changes the shape of the topological region and $Q_D(\phi, B)=-1$, various predicted universal features of the standard planar JJ model no longer hold. By comparing Figs.~\ref{fig:IcminLoc}(b) and \ref{fig:IcminLoc}(c), we see that, even at $\phi=\pi$, the location of the dip is outside of the topological region. Furthermore, this location does not coincide with that of the phase jump, nor with the location for the transition of the phase-unbiased system into the topological region,\cite{Dartiailh2021:PRL} as seen in Figs.~\ref{fig:BT_Oscillations}(b) and \ref{fig:BT_Oscillations}(c), red line.

With the recently renewed interest in nonreciprocal transport and diode effects in superconducting junctions,\cite{Amundsen2022:RMP,Nadeem2023:NRP} we note  that the 2019 report of the phase signature of the transition to topological superconductivity in planar JJs was also accompanied by the measured superconducting diode effect,\cite{Dartiailh2021:PRL,Amundsen2022:RMP} but was not discussed. The diode effect, while not considered in the standard model of planar JJs, is also inherent to these systems.\cite{Yokoyama2014:PRB} We parametrize the efficiency of the diode effect by the asymmetry of the critical currents\cite{Hou2023:PRL}
\begin{equation}
\eta \equiv \left(I_c^+-I_c^-\right)\,/\,\left(I_c^+ + I_c^-\right),
\label{eq:eta}
\end{equation}
shown along with the forward and reverse $\widetilde{I\,}^{\pm}_c$ for $W_S=128\;$nm in Figs.~\ref{fig:IcminLoc}(c), \ref{fig:IcminLoc}(d) for $W_N=160\;$nm, and in Figs.~\ref{fig:IcminLoc}(e)-\ref{fig:IcminLoc}(f). Red/blue curves retain our convention for $\alpha=5,15\;$eVnm. An observed trend here, that $\widetilde{I\,}^+_c$ and $\widetilde{I\,}^-_c$ will differ more for a larger $\alpha$ and larger $W_N$, is expected from the spin-dependent phase shift of electron and hole forming the Andreev bound states.\cite{Yokoyama2014:PRB} Away from the dip locations of $\widetilde{I\,}_c^\pm$, $\eta\sim 0.1$, while near the $\widetilde{I\,}_c^\pm$ $\eta$ becomes much larger, but also accompanied with only a small $I_c$.\cite{Pekerten2022:PRB,Lotfizadeh2024:CP} The sign change locations in $\eta$ correspond to either the minimized singlet or the triplet part of the finite-momentum pairing wavefunctions, which occur at different locations for narrow and wide JJs.\cite{Lotfizadeh2024:CP} These finite-size effects could guide future experiments on both superconducting diode effect and topological phase transitions in JJs.

Most of the work on topological JJs considered only Rashba SOC. While for certain crystallographic directions the spin texture from Dresselhaus SOC\cite{Zutic2004:RMP} recovers the one expected from only Rashba SOC in Eq.~(\ref{eq:SOC}), with the simultaneous presence of these two SOC contributions, present in many 2DEGs, the topological phase diagram can be strongly modified. In a situation from Fig.~\ref{fig:System}, where in-plane ${\bm B}$ is perpendicular to the current direction and for a narrow JJ with $\beta\neq0$, one can find that $\Delta_\mathrm{top}$ together with MBS protection are maximized for $\theta_c=\pi/4$ for an arbitrary ratio $\alpha/\beta$, i.e. an arbitrary $\theta_\mathrm{SO}$,\cite{Pekerten2022:PRB,Pakizer2021:PRR,Pakizer2021:PRB} recall Eq.~(6). We thus limit our results to $\theta_c=\pi/4$ and consider the cases where Rashba and Dresselhaus have equal strengths, $\theta_\textrm{SO} = \pi/4$, or the system has only Dresselhaus SOC, $\theta_\textrm{SO} = \pi/2$, while keeping constant the combined SOC magnitude, $\lambda_\mathrm{SO}$, defined in Eq.~(6).

\begin{figure}[t]
\centering
\includegraphics*[width=\columnwidth]{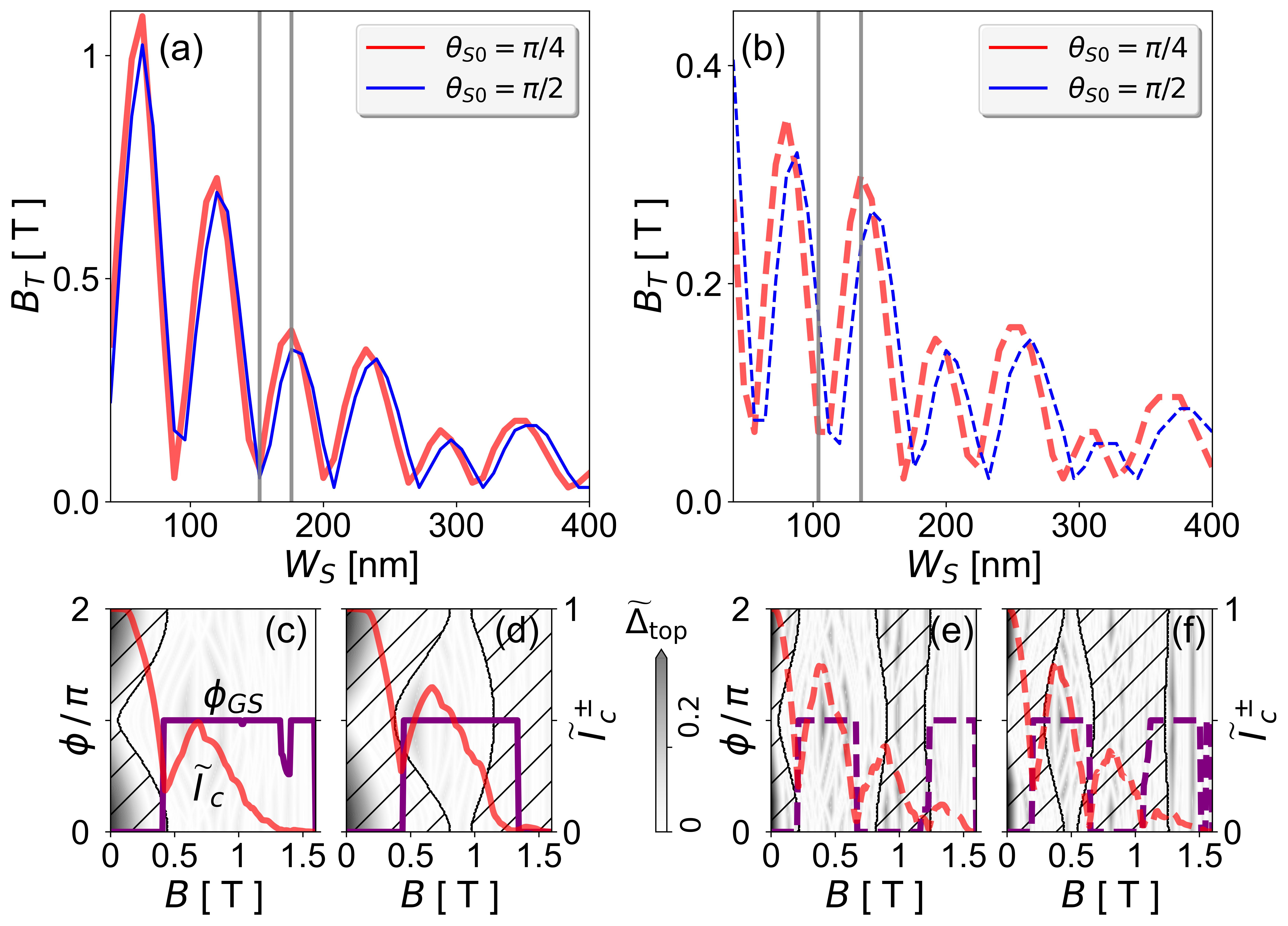}
\caption{Effects of Dresselhaus SOC; in all plots $\lambda_\mathrm{SO} = \sqrt{\alpha^2+\beta^2} = 15\;$meVnm and $\theta_c=\pi/4$, see Eq.~(2). Solid/dashed lines indicate $W_N$ as in Figs.~\ref{fig:BT_Oscillations},~\ref{fig:IcminLoc}. (a), (b) Oscillating $B_T(W_S)$; $\theta_\mathrm{SO}=\pi/4$ (red) [$\pi/2$, blue] corresponds to $\alpha=\beta$ ($\alpha =0$). (c)-(f) Topological phase diagrams with $Q_D(\phi, B)=1,-1$ (hatched, unhatched) and $\widetilde{\Delta}_\mathrm{top}(\phi, B)$ (grayscale), with $\phi_{GS}(B)$ (violet, right axis) and $\widetilde{I\,}_c(B)$ (red, left axis). (c), (d) correspond to a minimum, peak in $B_T(W_S)$ at $W_S=152,\,176\;$nm [gray lines in (a)]. (e), (f) correspond to a minimum, peak at $W_S=104,\,136\;$nm [gray lines in (b)].}
\label{fig:BT_Dresselhaus} 
\end{figure}

To consider the influence of Dresselhaus SOC, in Figs.~\ref{fig:BT_Dresselhaus}(a) and (b), we first revisit the oscillation of the topological transition field, $B_T(W_S)$, shown previously in Fig.~\ref{fig:BT_Oscillations}(a) for $\beta=0$. By choosing $\lambda_\mathrm{SO}=15\;$meVnm in Fig.~\ref{fig:BT_Dresselhaus}, we have a direct comparison with the red curves for  $\alpha=15\;$meVnm from Fig.~\ref{fig:BT_Oscillations} which show very similar decaying oscillations with $W_S$, while the results from Figs.~\ref{fig:BT_Dresselhaus}(a) and (b) again confirm that amplitude of these oscillations is suppressed with $W_N$. We also see that, it is not only the combined SOC magnitude, but also the relative contribution of Dresselhaus and Rashba  SOC which determines the oscillations in $B_T(W_S)$, shifted to the right for a larger $\beta$ ($\theta_\mathrm{SO} = \pi/2$) that becomes more pronounced with a wider $W_N$. As a result, while the oscillating trends in $B_T(W_S)$ are retained without and with Dresselhaus SOC, at a given $W_S$ and $W_N$, a change in the relative strength of the Rashba and Dresselhaus SOC can strongly alter the value of $B_T$ and thus also the topological phase diagram. For example, at $W_S=104\;$nm, in Figs.~\ref{fig:BT_Oscillations}(a) and \ref{fig:BT_Dresselhaus}(b) we can compare the three corresponding $B_T$ values with the same SOC strength $\lambda_{SO}=15\;$meVnm, but with various relative strengths of the Dresselhaus SOC. Using a pure Dresselhaus SOC as a reference, $B_T$ for a mixed Rashba and Dresselhaus SOC in Fig.~\ref{fig:BT_Dresselhaus}(b) differ by $\sim 0.1\;$T, while for a pure Rashba SOC in Fig.~\ref{fig:BT_Oscillations}(a), the difference in  is $\sim 0.08\;$T. Correspondingly, with a change in the relative strength of  Rashba and Dresselhaus SOC, the topological phase diagram and the onset of topological superconductivity can be noticeably different.

Another influence of Dresselhaus SOC can be seen in Fig.~\ref{fig:BT_Dresselhaus}(c) and \ref{fig:BT_Dresselhaus}(d) for the two chosen values of $W_S$, marked by gray lines in Figs.~\ref{fig:BT_Dresselhaus}(a) and \ref{fig:BT_Dresselhaus}(b) and matching the $W_S$ values in Fig.~\ref{fig:BT_Oscillations}. Several differences from $\beta=0$ and Figs.~\ref{fig:BT_Oscillations}(b) and \ref{fig:BT_Oscillations}(c) can be inferred, where the evolution of  $\widetilde{I\,}_c$ with $B$ was overall smoother and the jumps in $\phi_{GS}$ were less abrupt than seen in Fig.~\ref{fig:BT_Dresselhaus}(c) and \ref{fig:BT_Dresselhaus}(d). This is consistent with a shallower landscape for the free energy when $\beta \neq 0$, which supports more rapid changes with $B$, including the dip in the $\phi_{GS}$  in Fig.~\ref{fig:BT_Dresselhaus}(c). We note that the choice of ${\bm B}$ along the y axis in Fig.~\ref{fig:System}(a) maximized  ${\Delta}_\mathrm{top}(\phi, B)$ for $\beta=0$ and led to its reduction in Fig.~\ref{fig:BT_Dresselhaus}(c)-\ref{fig:BT_Dresselhaus}(f). With a larger $W_N=600\;$nm in Figs.~\ref{fig:BT_Dresselhaus}(e) and \ref{fig:BT_Dresselhaus}(f), there are even more abrupt changes in $\widetilde{I\,}_c$ and $\phi_{GS}$. While not shown in Fig.~\ref{fig:BT_Dresselhaus}, $\beta \neq 0$ again supports the superconducting diode effect, discussed in Fig.~\ref{fig:IcminLoc}.
 
Planar JJs provide a versatile platform for topological superconductivity as well as for superconducting spintronics,\cite{Linder2015:NP,Eschrig2015:RPP,Cai2023:AQT} through spin-triplet supercurrents and superconducting diode effect.\cite{Banerjee2014:NC,Amundsen2022:RMP,Nadeem2023:NRP} Even for a perfect interfacial transparency, we find a much richer behavior than what is expected from the common description of planar JJs. There are interesting future generalizations where our results could offer a helpful starting point. For example, in exploring the role of disorder in planar JJs,\cite{Stern2019:PRL,Kurilovich2023:NC} or more complex geometries,\cite{Kotetes2019:PRL,Pankratova2020:PRX,Schmitt2022:NL} it is important to include the relevant form of SOC and the finite-size effects. With a growing class of materials used in JJs, SOC that is linear in the wave vector may not be sufficient and the cubic SOC terms could become dominant.\cite{Samokhin2022:AP,Alidoust2021:PRB,Luethi2023:PRB,Tosato2023:CM} As a result, instead of the spin-triplet $p$-wave, the proximity-induced topological superconductivity could acquire the character of the spin-triplet $f$-wave.\cite{Alidoust2021:PRB}

The data that support the findings of this study are available from the corresponding author upon reasonable request.

\begin{acknowledgments} 
This work is supported by US ONR under awards MURI N000142212764, N000141712793, and NSF ECCS-2130845. Computational resources were provided by the UB Center for Computational Research.
\end{acknowledgments}

%Create the reference section using BibTeX:
%\bibliography{BP}

%merlin.mbs aipnum4-1.bst 2010-07-25 4.21a (PWD, AO, DPC) hacked
%Control: key (0)
%Control: author (8) initials jnrlst
%Control: editor formatted (1) identically to author
%Control: production of article title (0) allowed
%Control: page (1) range
%Control: year (1) truncated
%Control: production of eprint (0) enabled
%

\end{document}